\documentclass{emulateapj}

\newcommand{\mytilde}{\raise.17ex\hbox{$\scriptstyle\sim$}}
\newcommand{\brk}[1]{$[$#1$]$}
\newcommand{\kepler}{{\it Kepler}}
\newcommand{\corot}{{\it CoRoT}}

\newcommand{\mps}{m~s$^{-1}$}

\newcommand{\kps}{km~s$^{-1}$}

\newcommand{\gpcmthree}{g~cm$^{-3}$}
\newcommand{\microm}{$\mu$m}

\newcommand{\Rsun}{${\rm R_\odot}$}
\newcommand{\Mjup}{${\rm M_J}$}
\newcommand{\xonb}{XO-1b}
\newcommand{\xon}{XO-1}
\newcommand{\Rjup}{${\rm R_J}$}
\newcommand{\vMs}{1.027}		
\newcommand{\eMs}{0.06}
\newcommand{\vRs}{0.94}		
\newcommand{\eRs}{0.02}

\newcommand{\vrvK}{116} 
\newcommand{\ervK}{9}

\newcommand{\vjd}{2453887.74774}	
\newcommand{\ejd}{0.00011}	
\newcommand{\vjdOne}{2454506.56417}	
\newcommand{\ejdOne}{      0.00010}	
\newcommand{\vjdTwo}{2454518.38906}	
\newcommand{\ejdTwo}{      0.00017}	
\newcommand{\vap}{0.049}	
\newcommand{\eap}{0.001}

\newcommand{\vperiod}{3.94150685}	
\newcommand{\eperiod}{0.00000091}
\newcommand{\vMp}{0.92}
\newcommand{\eMp}{0.08}	
\newcommand{\vRp}{1.21}
\newcommand{\eRp}{0.03}
\newcommand{\vincl}{88.8}
\newcommand{\eincl}{0.2}
\newcommand{\vdur}{2.971}
\newcommand{\edur}{0.006}

\newcommand{\eUone}{0.07}
\newcommand{\vUtwo}{0.35}
\newcommand{\eUtwo}{0.08}

\newcommand{\vFeH}{0.02}
\newcommand{\eFeH}{0.08}
\newcommand{\vTeff}{5750}
\newcommand{\eTeff}{75}
\newcommand{\vgs}{4.50}
\newcommand{\egs}{0.01}
\newcommand{\vadr}{11.24}
\newcommand{\eadr}{0.09}
\newcommand{\vbimp}{0.23}
\newcommand{\ebimp}{0.04}
\newcommand{\vdens}{1.73}
\newcommand{\edens}{0.04}
\newcommand{\vsaf}{0.073}
\newcommand{\esaf}{0.006}
\newcommand{\vrpda}{0.0117}
\newcommand{\erpda}{0.00012}
\newcommand{\vrsda}{0.0890}
\newcommand{\ersda}{0.0007}
\newcommand{\vrsrp}{0.1007}
\newcommand{\ersrp}{0.0008}
\newcommand{\vrho}{0.1320}
\newcommand{\erho}{0.0005}
\newcommand{\vdensp}{0.64}
\newcommand{\edensp}{0.05}
\newcommand{\vlggp}{3.19}
\newcommand{\elggp}{0.03}
\newcommand{\ving}{0.366}
\newcommand{\eing}{0.007}

\slugcomment{Submitted to the Astrophysical Journal}

\received{April 29, 2010}
\begin{document}

\title{NICMOS Observations of the Transiting Hot Jupiter XO-1\lowercase{b}}

\author{
Christopher~J.~Burke\altaffilmark{1}
P.~R.~McCullough\altaffilmark{2},
L.~E.~Bergeron\altaffilmark{2},
Douglas~Long\altaffilmark{2},
Ronald~L.~Gilliland\altaffilmark{2},
Edmund~P.~Nelan\altaffilmark{2},
Christopher~M.~Johns-Krull\altaffilmark{3},
Jeff~A.~Valenti\altaffilmark{2},
Kenneth~A.~Janes\altaffilmark{4}
}

\email{cburke@cfa.harvard.edu}
\altaffiltext{1}{Harvard-Smithsonian Center for Astrophysics, 60 Garden St., Cambridge, MA 02138}
\altaffiltext{2}{Space Telescope Science Institute, 3700 San Martin Dr., Baltimore, MD 21218}
\altaffiltext{3}{Dept. of Physics and Astronomy, Rice University, 6100 Main Street, MS-108, Houston, TX 77005}
\altaffiltext{4}{Boston University, Astronomy Dept., 725 Commonwealth Ave.,Boston, MA 02215}

\begin{abstract}

We refine the physical parameters of the transiting hot Jupiter planet
\xonb\ and its stellar host \xon\ using HST NICMOS observations.
\xonb\ has a radius $R_{p}$=\vRp$\pm$\eRp\ \Rjup, and \xon\ has a radius $R_{\star}$=\vRs$\pm$\eRs\ \Rsun, where the uncertainty in the
mass of \xon\ dominates the uncertainty of $R_{p}$ and $R_{\star}$.
There are no significant differences in the \xon\ system properties
between these broad-band NIR observations and previous determinations
based upon ground-based optical observations.  We measure two transit
timings from these observations with 9~s and 15~s precision.  As a
residual to a linear ephemeris model, there is a 2.0 $\sigma$ timing
difference between the two HST visits that are separated by 3 transit
events (11.8 days).  These two transit timings and additional timings
from the literature are sufficient to rule out the presence of an Earth
mass planet orbiting in 2:1 mean motion resonance coplanar with \xonb.  We
identify and correct for poorly understood ``gain-like'' variations
present in NICMOS time series data.  This correction reduces the
effective noise in time series photometry by a factor of two, for the
case of \xon.

\end{abstract}

\keywords{planetary systems -- stars: individual (GSC 02041-01657)}

\section{Introduction}\label{sec:intro}

\xonb\ is a transiting hot Jupiter planet orbiting a Solar-type star
\citep{MCC06}.  Subsequently, \citet{HOL06} refined
the parameters of the star, \xon, and planet, \xonb, with ground based
optical photometry \citep{TOR08,SOU08}.  From a theoretical standpoint
\xonb\ has the properties of a ``normal'' extrasolar planet, in the
sense that the measured mass, $M_{p}$, and radius, $R_{p}$, of \xonb\
agree with the theoretical expectations after taking into account the
stellar insolation, and \xonb\ likely does not contain a substantial
amounts of heavy elements \citep{BURR07}.  Spitzer Space Telescope
observations with IRAC of \xonb\ at secondary eclipse detected the
presence of a temperature inversion in the outer layers (1 mbar) of
the planetary temperature-pressure (T-P) profile \citep{MAC08}.  The
wavelength dependence of the IRAC measurements provides a valuable
constraint on the day-side T-P profile and molecular abundances.

Measuring the absorption depth when the planet transits its stellar
host as a function of wavelength, here referred to as transmission
spectroscopy, provides additional constraints on the planet's
atmosphere.  The change in transit depth results from opacity
variations as a function of wavelength in the planet's atmosphere and
probes the T-P profile in the outermost ($<$ 1 mbar) layers of the
planet's atmosphere at the terminator \citep{FOR10}.  To extend
transit observations of \xonb\ to the NIR, we present high cadence HST
NICMOS observations of \xonb.  The observations were obtained with the
G141 grism, which covers the strong absorption features due to
H$_{2}$O, CO$_{2}$, and CH$_{4}$.  In this work, we present new
constraints on the \xon\ system by summing the flux over the entire
grism spectrum.  The NICMOS broad-band time series provide a precise
measurement of $R_{p}$, stellar density, $\rho_{\star}$, and transit
timings for which to compare to previous determinations in the
optical.  \citet{TIN10} present a complementary study to this one
using the same NICMOS observations as analyzed here by enhancing the
wavelength resolution at the sacrifice of signal to noise.
\citet{TIN10} examine the relative change in transit depth across the
G141 grism, and they detect the presence of H$_{2}$O in the atmosphere
of \xonb.

Currently, NICMOS grism observations for 9 transiting extrasolar
planets have used $\gtrsim$120 HST orbits.  Previous analyses of grism
broad-band time series with NICMOS have been limited to a factor of
2-3 times worse than the Poisson expectation (HD 209458b -
\citet{GIL03}, \citet{SWA09}; GJ 436b - \citet{PON09}; HD 149026b
\citet{CAR09B}).  In this work, we identify and correct for
systematics that improve the relative flux precision for broad-band
time series to the expected level.  The technical details of this work
will augment the substantial archival NICMOS grism observations and
future NICMOS grism observations of transiting extrasolar planets.

The NICMOS observations are described in \S~\ref{sec:nicobs}.  The
technical details of the data analysis, improvements in the treatment
of systematics for NICMOS grism observations, and a discussion of the
expected photometric noise level are given in \S~\ref{sec:lc}.  The
\xon\ system properties are discussed in \S~\ref{sec:res}, and we discuss the results in \S~\ref{sec:disc}.

\section{Observations}\label{sec:nicobs}

The NICMOS grism observations were designed to gather high cadence
time series of the bright, V=11.2, star \xon\ during transits of the
planet \xonb\ (HST program 10998).  Gaps in the time series due to
Earth occultation necessitate piecing together observations from more
than one transit.  HST observed a transit event on Feb.\ 10, 2008 UT
(2454506.6 JD) and again three transits later on
Feb.\ 21, 2008 UT (2454518.4 JD).  Each visit consists of five HST
orbits.

\begin{figure}
\plotone{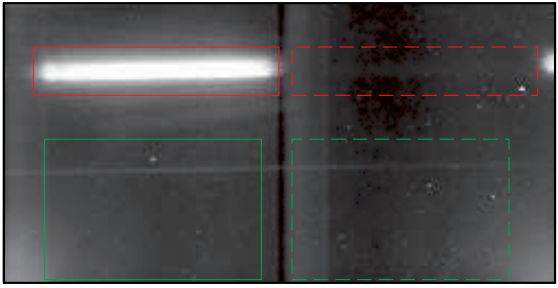}
\caption{
Lower two quadrants of the NIC3 camera showing the photometric and background
apertures for the target {\it solid} and blank {\it dashed} region,
left and right respectively.  The image is based upon averaging all
images in a single orbit.  In addition to the first-order spectrum in
the upper left, the zeroth-order spectrum is visible in the upper
right.  The readout amplifier glow is apparent in the lower corners
since we analyze raw data rather than the CALNICA pipeline calibrated
images, which have the amplifier glow removed.
\label{fig:img}}
\end{figure}

The first-order spectrum of \xon\ from the G141 grism ($1.1\leq
\lambda \leq 1.9$ \microm) is positioned on the lower left quadrant
(quad 1) of the NIC3 NICMOS detector, and it does not cross any
amplifier boundaries (see Figure~\ref{fig:img}).  The zeroth-order
spectrum lies on the lower right quadrant (quad 2), and it enables
accurate tracking of the spectrum's position.  No other stars
contribute significant flux in the NIC3 field of view.  The Pupil
Alignment Mechanism was set at $-0.53$~mm to defocus the Point Spread
Function (PSF).  Defocusing spreads the light over more pixels which
improves operational efficiency by delaying saturation and improves
precision by averaging nonuniform pixel response \citep{XU03}.  The
detector is read out in MULTIACCUM mode with the STEP8 sequence and
NSAMP=11, resulting in an overall exposure time 39.953~s for a single
MULTIACCUM exposure.  Including overheads, the exposure cadence is
49~s between MULTIACCUM exposures.  There are $\sim$57 exposures per
HST orbit.  A single direct image in the F166N filter at the beginning
of each HST visit provides the reference position of the target for
determining the wavelength calibration.

We reduced the data using both custom procedures and
publicly-available procedures written in IDL.  The reduction begins
from the raw science file (\_raw) rather than the calibrated science
file (\_cal).  In summary, the procedure for reducing an image
consists of starting from the last-read minus zeroth-read image of a
MULTIACCUM image, applying a wavelength dependent flat field, and then
correcting bad pixels through bicubic spline interpolation.  These
data reduction steps are described next, and the resulting images are
used to perform broad-band photometry in \S~\ref{sec:lc}.

\begin{figure}
\plotone{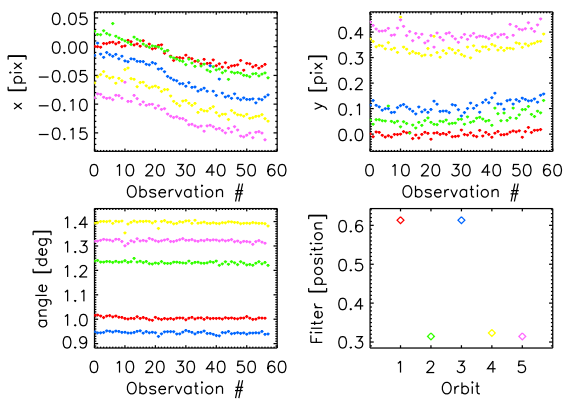}
\caption{{\it Top Panels:}
Evolution of the dispersion relation's reference x and y position
throughout the five orbits of the first visit as a function of image
number within an orbit.  Each color represents an individual orbit as
given in the key in the bottom right panel.  {\it Bottom Left:}
Spectrum slope as a function of observation number within an
orbit.  {\it Bottom Right:} Filter wheel telemetry for each orbit and
color key for each orbit within the first visit.\label{fig:ext1}}
\end{figure}

\begin{figure}
\plotone{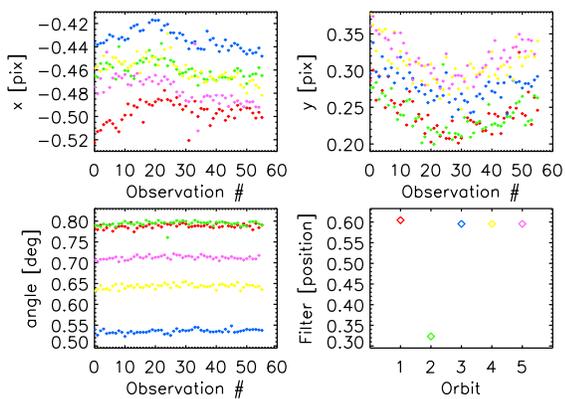}
\caption{
Same as Figure~\ref{fig:ext1}, but for the second visit.\label{fig:ext2}}
\end{figure}

To determine the wavelength dependent flat field and to correct for
systematic trends in the photometric time series (see
\S~\ref{sec:lc}), we map the position of the spectrum on the detector.
To measure the spectrum location, we fit a Gaussian to calculate the
centroid of the spatial profile for 5 pixel intervals along the
dispersion axis.  A linear fit to the centroids versus position along
the dispersion axis yields the slope of the spectrum on the detector.
The lower left panel in Figures~\ref{fig:ext1}~\&~\ref{fig:ext2} show
the spectrum slope as a function of image number within an orbit for
the first and second visit, respectively.  Each color represents an
orbit, and the color code to identify an orbit is given in the lower
right panel of the figure.  The spectrum slope variation within an
orbit is much smaller than the change between orbits.  We measure
relative shifts in x and y pixel coordinates of the spectrum by cross
correlation of the zeroth-order spectrum with respect to a reference
image (11-th image in the second orbit of the first visit).  The
zeroth-order position shifts are projected to the first-order
spectrum's dispersion relation position.  The projection, $\delta
x=-\Delta\cos{\alpha}$ and $\delta y=-\Delta \sin{\alpha}$, takes into
account the spectrum slope, $\alpha$, and the pixel distance between
the zeroth-order light and dispersion relation's reference position,
$\Delta=167.93$~pix.  The top panels in Figure~\ref{fig:ext1} and
Figure~\ref{fig:ext2} show the relative x and y pixel coordinate
shifts of the dispersion relation's reference position as a function
of the image number within an orbit.  The dispersion reference
position is repositioned within 0.5~pixel between visits and typically
0.2~pixel within a visit.

We follow the procedure outlined in \citet{GIL03} to determine the
wavelength dependent flat field, and the wavelength of each pixel is
determined using the dispersion relation given in \citet{PIR09}.  The
centroid of the direct image taken at the beginning of each visit
determines the reference position of the dispersion relation.  A flat
field is calculated for each image taking into account the reference
position shifts and the spectrum slope variations.  The individual
flat fields are averaged over an orbit, and the resulting orbit
averaged flat field is normalized on quad 1 where the
first-order spectrum is located.  The orbit averaged flat field is applied
to all images within that orbit.

We use four methods to identify the dark, warm, and cosmic ray
impacted pixels.  We begin with the standard bad pixel mask for the
NIC3 camera \citep{SOS02}.  Second, we identify pixels that
significantly vary beyond their empirically determined sample standard
deviation throughout an orbit.  Third, warm pixels ($>$ 100 DN) in the
region for determining the background level are identified for
correction.  Fourth, pixels that deviate by $>$~3$\sigma$ from the
Poisson and read noise expectation in both forward and backward
differences are identified for correction.  On average, each image has
177 pixels (1.1\%) for correction on quad 1.  We replace affected
pixels by the value of a bicubic spline fitted to surrounding pixels.

In addition to the spectrum position and slope, we make use of the
G141 filter wheel telemetry to correct for systematic trends in the
photometric time series.  The filter wheel telemetry indicates that
the G141 grism position does not return to the same state in between
orbits.  The filter wheel telemetry for each visit (Chris Long,
private communication) is given as a function of orbit in the lower
right panel of Figures~\ref{fig:ext1}~\&~\ref{fig:ext2}.  There are
two preferred telemetry states of the G141 grism ($\sim$0.6 and
$\sim$0.3 fractional telemetry position), and within each of these
telemetry states the telemetry do not exactly repeat but have small
variations.  In the first visit (Figure~\ref{fig:ext1}), the filter
wheel telemetry anti-correlates with the spectrum slope; The filter
wheel telemetry position of 0.6 (orbits 1 \& 3) correspond to low
spectrum slope, and the filter wheel telemetry position of 0.3 (orbits
2, 4, \& 5) corresponds to the higher spectrum slope.  In the second
visit, the correspondence of filter wheel telemetry and spectrum slope
is not as clear.

\section{Broad-Band Light Curve Analysis}\label{sec:lc}

This section describes the process to extract the broad-band
photometric signal from the calibrated images by summing the source flux
over the entire first-order spectrum.  We describe newly identified
systematics of the NIC3 camera and the procedure we follow to
simultaneously correct the systematics and fit for the transiting
planet model.  The result is an improved estimate for the physical
properties of the planet \xonb\ and star the \xon.

\subsection{Photometry and Expected Noise}
We define the source flux to be the sum of counts in a rectangular
region enclosing the first-order light minus an estimate of the
background flux contribution.  The positional shifts of the
first-order spectrum ($<1$ pix) are small enough to adopt the same
photometric aperture and background region for all visits and orbits.  The
photometric aperture is a 113x23~pixel region, and the background estimate
aperture is 100x65~pixel region located 25 pixels below the
photometric aperture.  The solid rectangular outlines in
Figure~\ref{fig:img} illustrate the photometric and background regions, top
left and bottom left, respectively.  These regions were chosen to
minimize flux variations in the extracted photometry after correcting
for systematic effects.

\begin{figure}
\plotone{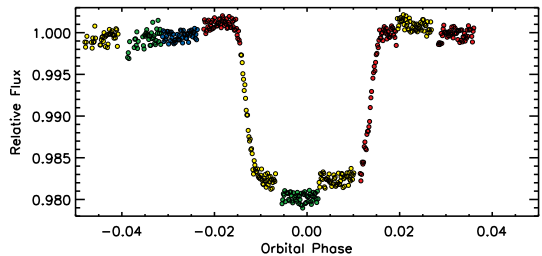}
\caption{
Raw relative observed counts for \xon\ phased at the \xonb\ orbital
period.  Data is shown from the first HST visit ({\it green \& red
points}) and the second HST visit ({\it yellow \& blue points}).
Within a visit, orbits that share the same color share the same filter
wheel position.\label{fig:rawlc}}
\end{figure}

The first-order summed flux from \xon\ is shown in
Figure~\ref{fig:rawlc}, where the flux relative to the average out of
transit flux is shown phased at the orbital period of \xonb.  Each HST
visit is normalized separately.  There is a 0.6\% relative flux
difference between the two HST visits and smaller, but significant
0.2\% relative flux differences between orbits within the same visit.
The upper left panel of Figure~\ref{fig:7cor} shows in detail the
in-transit orbit closest to orbital phase, $\phi=0.0$.  For those
data, the empirical rms noise is $\sigma=555$ ppm per exposure.  The
expected noise from Poisson noise alone is $\sigma_{\rm poi}=180$ ppm
per exposure.  An additional source of noise arises due to the
systematic uncertainty in determining the background level and read
noise.  To empirically determine the contribution of read noise and
background level subtraction to the photometric precision, we perform
the aperture photometry on a blank region of quad 2.  The target and
background regions are the same size as for the first-order spectrum
photometry as illustrated by the dashed line regions in
Figure~\ref{fig:img}.  The expected noise in the blank region
photometry taking into account the Poisson noise, read noise, and the
number of pixels in the background and photometry regions is 2300
e$^{-}$ rms.  We measure 5400 e$^{-}$ rms per exposure variation in
the blank region photometry, indicating the presence of non-Gaussian
noise in the background.  The empirically determined read noise and
systematic uncertainty in the determining the background level
contributes $\sigma_{\rm back}$=170 ppm noise relative to the
3.1$\times 10^7$ e$^{-}$ counts from \xon\ per exposure.  Adding
$\sigma_{\rm back}$ in quadrature with $\sigma_{\rm poi}$ results in
the expected relative noise for photometry of $\sigma_{\rm tot}=250$
ppm per exposure.  The expected noise is 2.2 times smaller than the
empirically determined noise within an orbit.  In addition to the
higher than expected noise, the residuals qualitatively are
distributed uniformly between limits and are non-Gaussian in
appearance.  This broadened distribution of residuals has been noted
previously \citep[see Figure 3 of ][]{CAR09B}.

\subsection{Correcting for Gain-like Variations}
The non-Gaussian residuals in the photometric time series correlate
with 7 preferred states imparted by the detector electronics that
result in gain-like variations.  The 7 states originate from the
detector's switching low voltage power supply, which has its clock set
to a frequency 7 times lower than the master clock of the timing
pattern generator (Bergeron, in preparation).  The 7 states were first
identified as additional noise in the detector temperature estimates
from the bias level \citep{PIR09b}.  In this study, we show that the 7
states also impact photometric time series with NICMOS.  Which of the
7 states an exposure belongs to is measurable from the zeroth-read
image of a MULTIACCUM image.  The zeroth-read image is obtained
immediately following a reset of the detector array and it represents
the bias level of the MULTIACCUM image.  For a more accurate measure
of the relative bias level, the zeroth-read image from the first image
within an orbit is subtracted, removing any bias structure and faint
signal of the spectrum recorded since the array reset.  The mean
zeroth-read bias level of quad 1 versus the mean zeroth-read bias
level of quad 2 for images within an orbit is shown in the upper right
panel of Figure~\ref{fig:7cor}.  The images in this plane form 7
distinct groups.  The lower left panel of Figure~\ref{fig:7cor} shows
the same relative photometry as in the panel above it, but the
measurements that share the same readout state are connected by a
line.  The color coding is the same as in the upper right panel of
Figure~\ref{fig:7cor}, where the 7 states are identified.
Measurements in the same state systematically over or underestimate
the relative flux.  We correct for the 7 state effect by removing the
average relative flux offset of the state from the relative flux
offset averaged over all measurements in an orbit.  The result of 7
state correction is shown in the lower right panel of
Figure~\ref{fig:7cor}.  The rms of the residuals is reduced from 550
ppm (before) to 288 ppm (after), a factor of 1.9 improvement.  To
properly account for the 7 states during times when the flux level is
varying rapidly (i.e., during ingress and egress), the 7 state
correction is determined after removing the transit model (see below).

\subsection{Model Fit to Data}

The remaining flux time series, $F_{obs}$, are fit to the following
model, \begin{equation} F_{mod}=F_{o}\times \Psi \times \Phi
(P,T_{o},M_{\star},R_{\star},\delta,\tau_{I-IV},t_{1},t_{2},u_{1},u_{2}),\label{eq:mod}
\end{equation} where $F_{o}\times\Psi$ is the correction for the
observed flux variability due to instrumental effects and intrinsic
stellar variability between visits (see below) and $\Phi$ represents
the relative flux variation due to the transiting planet.  The orbital
period, $P=3.941502$ d (Gary, B., private communication), and
ephemeris transit midpoint, $T_{o}=2453887.74679$ \citep[HJD;
][]{HOL06}, are held fixed, but each visit is allowed to have its own
transit midpoint offset ($t_{1}$ \& $t_{2}$) from the fixed ephemeris.
This initial ephemeris is based upon a compilation of professional and
amateur observations of XO-1.  The longer time baseline provided by
including amateur observations allowed a more precise estimate of the
ephemeris than previously published ephemerides.  We adopt a quadratic
limb darkened ($u_{1}$\& $u_{2}$) transit model as given by
\citet{MAN02}, and use a Gaussian prior on the stellar mass,
$M_{\star}=$\vMs$\pm$\eMs\ $M_{\odot}$ \citep{TOR08}.  The remaining
parameters in $\Phi$ are stellar radius, $R_{\star}$, planet radius to
stellar radius ratio, $\delta=R_{p}/R_{\star}$, and first-to-fourth
contact transit duration, $\tau_{I-IV}$.  From the constraints placed
by radial velocity measurements of the \xonb\ orbit \citep{MCC06}, we
assume zero eccentricity.  The parameter estimates and their
uncertainty are determined following the MCMC procedure outlined in
\citet{BUR07}.  The prior adopted for $R_{\star}$, $\delta$, and
$\tau_{I-IV}$ results in a uniform prior on $R_{\star}$, $R_{p}$, and
orbital inclination \citep{BUR07}.  The priors for the other
parameters are uniform with upper and lower limits well beyond values
constrained by the data.  The exception to this is limb darkening
parameters, which are physically constrained as described in
\citet{BUR07}.  In particular, we require that the highest surface
brightness be located at the center of the disk (i.e., $u_{1}\geq 0$).
This is consistent with theoretical H-band limb darkening parameters
($u_{1}=0.016$ \& $u_{2}=0.441$) for \xon\ \citep{CLA00}.  The time
for each observation is taken as the midpoint of the exposure, and is
determined from the average of the EXPSTART and EXPEND header
keywords.  The resulting modified Julian date is converted to
barycentric Julian date (BJD) on the Terrestrial Time system (TT) using the barycen IDL
routine\footnote{http://astro.uni-tuebingen.de/software/idl/aitlib/astro/barycen.html
; Input ephemeris file from http://www.physics.wisc.edu/\mytilde
craigm/idl/down/JPLEPH.405} \citep{EAS10}.  The calculated BJD times are $\sim$1.3~s
earlier than heliocentric Julian date times.

The first orbit of a visit exhibits a systematic ramp up in counts
over most of the orbit, which likely results from telescope or
instrument settling \citep{GIL03}.  Subsequent orbits within a visit
also have a ramp up effect, but the timescale is much shorter with
only the first few measurements impacted.  To avoid modeling the
ramp-up effect, we discard the entire first orbit of a visit and the
first four exposures of each orbit.

To decorrelate against external parameters to correct for the
remaining correlated measurements, through trial and error, we find
that the remaining residuals correlate with the external parameters:
the spectrum's dispersion relation positions, $x$ \& $y$, and the
spectrum slope $\alpha$ (Figures~\ref{fig:ext1}~\&~\ref{fig:ext2}).
Thus,
\begin{equation}
F_{o}\times\Psi=F_{o}\times (1+c_{0}+c_{1}x+c_{2}x^2+c_{3}y+c_{4}\alpha ),
\label{eq:mod2}
\end{equation}
where $c_{0},...,c_{4}$ are the free linear decorrelation coefficients and
$F_{o}$ is arbitrary and fixed (to avoid the degeneracy with $c_{0}$)
at the average out-of-transit flux of both visits.  We empirically
find that a successful correction for correlated measurements using
the terms in Equation~\ref{eq:mod2} requires treating data from the
two filter wheel states separately (see \S~\ref{sec:nicobs} for a
discussion of the filter wheel states).  In Figure~\ref{fig:rawlc},
the green and red points represent data obtained in the first HST
visit, and the yellow and blue points represent data obtained in the
the second HST visit.  In each visit, the orbits are assigned to the
two preferred states of the filter wheel positioning (see lower right
panel of Figure~\ref{fig:ext1}~\&~\ref{fig:ext2}).  Each visit and
filter wheel state has its own set of external parameter decorrelation
coefficients, $c_{0},...,c_{4}$. Thus, in Figure~\ref{fig:rawlc}, the
orbits that share a color share the same decorrelation coefficients,
however as noted previously, the first orbit for each visit is not
included in the MCMC analysis.

There are 5 decorrelation coefficients for each of the 4 visit/filter
wheel combinations, resulting in 20 free parameters in $\Psi$.
Including the 8 free parameters specifying the physical parameters of
the system, the overall model has a total of 28 free parameters.
Before calculating the likelihood of a model for a given set of 28
parameters, the correction for the 7 readout states is applied.  The 7
readout states are determined for each orbit independently from the
residuals of the model, $F_{obs}-F_{mod}$.  The
7 state correction removes the average flux residual for each state
with respect to the average residual across all measurements within an
orbit.  Thus, the 7 state correction does not change the overall flux
level of an orbit.  The likelihood is modeled as independent Gaussian
residuals with uncertainty, $\sigma=250$ ppm.

The parameters and their uncertainties are based upon an overall MCMC chain
 of 10$^6$ steps after a burn-in period.  The scale of the
proposal steps in each parameter are set using an automated iterative
algorithm of proposal step size adjustments until the acceptance
fraction, $0.2<f<0.3$, is reached.  After an initial burn-in period to
determine the proposal steps, the proposal distributions are finalized,
and the results are based upon the remaining steps.  The longest auto
correlation length amongst the parameters is 600 steps.  Initial tests
with a parallel tempering (PT) MCMC algorithm \citep{GRE05} with 7
parallel chains, did not show any evidence for multimodality amongst
the 28 free parameters.  Thus, the more time-consuming PT MCMC was not
needed to explore the parameter space and helps verify that the single
chain reliably explored the parameter space.

\section{Results}\label{sec:res}

The resulting transit light curve, residuals, and corrections are
shown in Figure~\ref{fig:finlc}.  The top panel shows $F_{obs}/\Psi$,
the relative flux of the observations after dividing out the best fit
(in a $\chi^2$ sense) decorrelation function with correction for the 7
states.  The color coding of the points is the same as in
Figure~\ref{fig:rawlc}.  The middle panel of Figure~\ref{fig:finlc}
shows the residual relative flux of the observations from the complete
model.  The resulting rms residual, $\sigma_{\rm rms}=240$ ppm, is
slightly less than the expected uncertainty, $\sigma_{\rm tot}=250$
ppm (see \S~\ref{sec:lc}).  We presume the latter results from a
slight overestimate in the contribution of $\sigma_{\rm back}$ to
$\sigma_{\rm tot}$.  The lower panel of Figure~\ref{fig:finlc} shows
the relative flux correction due to $\Psi$.  The peak to trough
relative flux variation in $\Psi$ reaches $0.16\%$ for orbits within
the same visit and filter wheel position.  For display purposes,
$\Psi$ is normalized to its average value for each visit/filter wheel
combination.  Within a visit, the external parameter decorrelation
normalization, $c_{0}$, varies by $0.12\%$ in relative flux.  In the
12 days between the two visits, the observed flux of \xon\ varied by
$0.6\%$ due to intrinsic variability of the star or the instrument,
but we cannot distinguish between these two possibilities from these
data.  The minimum $\chi^2=362.8$ with 362 degrees of freedom
indicates the model is an acceptable fit to the data.

Tables~\ref{tab:planet}~\&~\ref{tab:star} show the resulting parameter
estimates and their uncertainty for \xonb\ and \xon\ along with
previous determinations from the literature, respectively.  The
parameter estimates come from the median of MCMC samples, and the
uncertainties inscribe 68.3\% of the MCMC samples.  The uncertainties
include the impact of the assumed prior on $M_{\star}=$\vMs$\pm$\eMs.
This affects the physical properties of the system that are not
directly constrained by the light curve (e.g., $R_{\star}$ \&
$R_{p}$).

\subsection{Light Curve Quality}
The light curve quality is high enough that the uncertainty in
$R_{\star}$ and $R_{p}$ is dominated by the uncertainty in
$M_{\star}$.  A solution assuming fixed $M_{\star}$ results in
$\sigma_{\star}=0.008$ \Rsun\ and $\sigma_{p}=0.013$ \Rjup\
uncertainty in $R_{\star}$ and $R_{p}$, respectively.
Table~\ref{tab:errors} illustrates the sensitivity of the parameters
that are directly constrained by the light curve (i.e., independent of
$M_{\star}$) to fixing the limb darkening parameters at their
theoretical expectation.  We adopt the H-band values ($u_{1}=0.016$ \&
$u_{2}=0.441$) from \citet{CLA00}.  Fixing the limb darkening
parameters results in a $\Delta \chi^2=1.3$ worse fit, a
non-significant difference in the quality of fit.  However, the
estimate for $R_{p}/R_{\star}$ is 1$\sigma$ smaller and the
uncertainty is 40\% smaller.  Fixing the limb darkening parameters can
lead to more precise model fits, but in the case of high quality data,
it may result in lower accuracy when the stellar brightness profile
differs from the theoretical expectation \citep{SOU08,CLA09}.  Also
shown in Table~\ref{tab:errors} is the precision possible if the
observations were free from systematics.  The results were obtained by
fixing the decorrelation coefficients, $c_{0},...,c_{4}$, at the
values that minimize $\chi^2$.  The need to correct for systematics
reduces the precision in $R_{p}/R_{\star}$ by a factor of 5 and
reduces the precision in the transit midpoint by a factor of 1.2.

The results shown in Table~\ref{tab:errors} emphasize the difficulty
in judging the light curve quality based upon comparing the
resulting uncertainty in the model parameter estimates alone.
Assuming a more precise estimate of $M_{\star}$ (or equivalently
$R_{\star}$), fixing the limb darkening parameters, or assuming fixed
decorrelation coefficients for in-transit data based upon their
solution from out-of-transit data, will all result in higher precision
for the system parameters, but such procedures may result in less
accuracy.  Comparing the quality of light curves should not be done
out of context.  For example, with comparable quality data, the
precision for a transit model analysis using fixed limb darkening
coefficients should not be directly compared to the precision that
results from a transit modeling analysis that allows the limb
darkening coefficients to vary.

\subsection{Understanding NICMOS Systematics}
Recently, \citet{TIN10} analyzed the same observations of \xon\ to
measure the transmission spectrum of the atmosphere of \xonb.
Although their analysis focused on individual spectral channels across
the G141 grism, their independent analysis techniques share some
similarities with the approach adopted in this study.  The 7 state
correction described in this study is closely related to their
correction for ``channel-to-channel'' correlations \citep{SWA08}.  The
gain-like variations associated with the 7 states, which appear to be
wavelength independent, would coherently affect the relative flux
residuals averaged over all wavelength channels, and be removed through
the ``channel-to-channel'' corrections.  The ``channel-to-channel''
correction of \citet{SWA08} cannot be applied to the broad-band
photometry, since by definition the broad-band photometry has a single
channel.  The study by \citet{TIN10} analyzed only the second HST
visit to \xon\ (yellow and blue points in Figure~\ref{fig:rawlc}), as
the first visit (green and red points in Figure~\ref{fig:rawlc}) was
deemed too photometrically unstable.  By treating orbits in different
filter wheel positions separately, we are able to provide reliable
model fits (reduced $\chi^{2}\sim 1$) using data from both HST visits
to constrain the properties of the \xon\ system.  Analysis of more
NICMOS observations are warranted, but it indicates our methodology
enables a coherent procedure to help analyze NICMOS datasets of
transiting planets that were previously thought to be too
photometrically unstable to provide useful results.

\subsection{Impact of Stellar Spots}
The light curve analysis assumes the stellar surface is described by
the limb darkening function.  However, the presence of dark spots or
bright faculae on the surface of the star violate this assumption
\citep[e.g., ][]{PON08}. The discovery photometry (0.8\% precision)
and slow rotation ($v\sin i < 3$\kps ) indicate that \xon\ is not an
active star \citep{MCC06}.  There is a 0.6\% apparent flux difference
in the measured counts between HST visits; \xon\ appears brighter
during visit 2.  The photometric stability of the NICMOS cameras is of
order $\sim 1$\% \citep{THA09}, and orbit to orbit differences of
0.2\% in flux within a single visit are present.  Thus, it is not
clear whether the flux difference between the first and second visit
is due to intrinsic variability in
\xon\ or due to instrumental photometric instability.  If the
$\eta=0.006$ decrease in flux results from the appearance of an
unocculted dark spot, then the $R_{p}/R_{\star}$ as measured for the
transit when the dark spot is present needs to be reduced by a factor
of $\sqrt{1-\eta}=0.997$ to compare to the measured $R_{p}/R_{\star}$
when the spot was absent (i.e., the transit depth is deeper when more
unocculted dark spots are present).  The expected 0.25\% change in
$R_{p}/R_{\star}$ between visits due to the presence of a hypothetical
dark spot, is smaller than our precision with which we currently
measure $R_{p}/R_{\star}$.  To verify this expectation, we added a
free parameter to our model allowing each visit to have its own
$R_{p}/R_{\star}$.  The resulting model negligibly improves the fit,
$\Delta \chi^2<1$, however $\Delta R_{p}/R_{\star}=-0.0003\pm0.0009$,
or 0.23\% lower $R_{p}/R_{\star}$ for the second visit.  This agrees
in size and direction as the expectation, however it is within the
statistical uncertainty, and thus we don't formally adjust the results
in Table~\ref{tab:planet}.

\subsection{Impact of Correlated Data}
\begin{figure}
\plotone{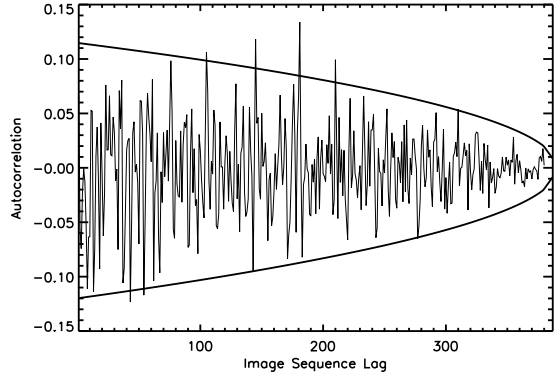}
\caption{
Sample autocorrelation of the model fit residuals as a function of image sequence lag.  Upper and lower smooth curves indicate the 3$\sigma$ limits for the null hypothesis that the residuals are independent for the sample autocorrelation statistic.\label{fig:autocor}}
\end{figure}

The likelihood used in the Bayesian posterior assumes the residuals
are independent and can result in underestimated parameter
uncertainties if the residuals are not independent
\citep{PON06,CAR09A}.  We analyze the sample autocorrelation function
\citep{BOX08} to verify that the correction for systematics removes
the temporal correlations present in the raw data, and that our use of
the likelihood that assumes independence is valid
(Figure~\ref{fig:autocor}).  To reliably determine the sample
autocorrelation especially for large lags, the sample autocorrelation
is calculated on the vector of residuals that are ordered by image
number rather than time.  The smooth upper and lower curves in
Figure~\ref{fig:autocor} shows the 3$\sigma$ limits as to the
expectation of the sample autocorrelation statistic if the underlying
population of residuals are independent \citep{KAN10}.  The sample
autocorrelation quickly dies off by lag 1, $a_{1}=-0.067$, but a few
of the lags spike above the 3$\sigma$ expectation.

We further explore the assumption of independence in the likelihood
through Monte Carlo simulation of the raw data and subsequent
retrieval of the system parameters.  We generate a red noise vector
that has the same variance and autocorrelation function as the model
residuals.  To ensure the sample autocorrelation function does not
underestimate the true autocorrelation function, the simulated red
noise has an autocorrelation for all lags ($\geq 1$) 1.5 times the
sample autocorrelation shown in Figure~\ref{fig:autocor}.  In addition
to the transit model with red noise, we add correlations with external
parameters and 7 state offsets.  By comparing the scatter of parameter
estimates derived from a $\chi^2$ minimization of the simulated light
curves and compare to the uncertainty estimates of an MCMC analysis of
the actual data for fixed stellar mass, we find the uncertainties
agree within 5\% for all the transit model parameters except for the
uncertainty in R$_{\star}$, which is 18\% larger.  However,
the larger uncertainty in R$_{\star}$ is negligible compared to the
much larger uncertainty in R$_{\star}$ that results from the
uncertainty in M$_{\star}$.  These tests show that the use of a
likelihood assuming independence does not result in a significant
underestimate in the parameter uncertainties.

\subsection{Transit Timing Variations}
\begin{figure}
\plotone{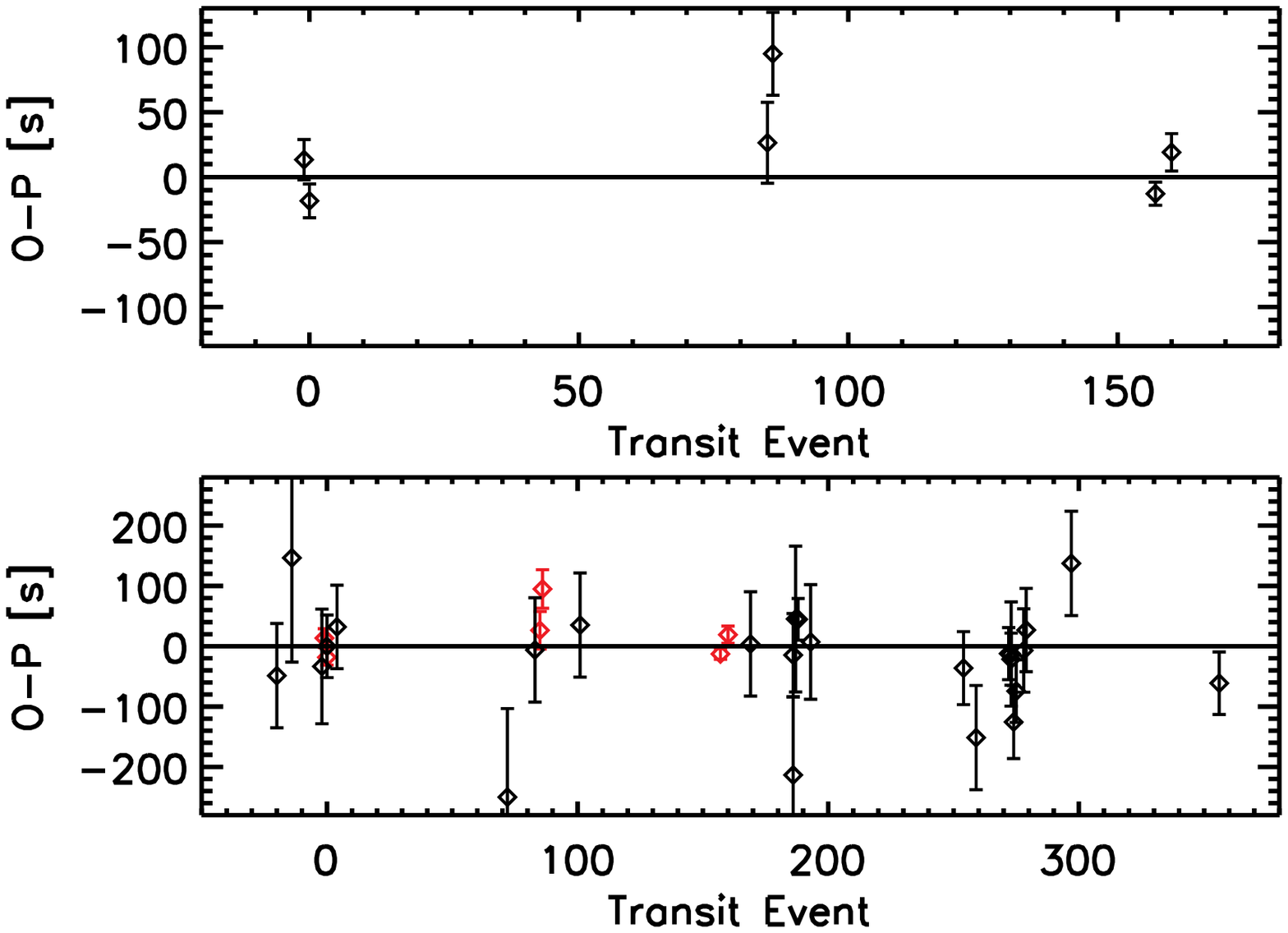}
\caption{{\it Top:} Deviations of high precision transit timings of \xonb\ from the literature and new timings from this work from a linear ephemeris model.  From left to right, measurements with Keplercam on the FLWO 1.2m \citep{HOL06}, SOFI on the NTT and ISAAC on UT1 \citep{CAC09}, and results from the two HST visits reported in this work.  The ephemeris of \xonb\ is calculated from these transit timings.  {\it Bottom:} Deviations of additional transit timing measurements of \xonb\ from the ephemeris derived in this work ({\it black points}).  The timing residuals from the top panel are reproduced in this panel ({\it red points}).\label{fig:ttv1}}
\end{figure}

The exposure timing information in the NICMOS image headers is on a
UTC system, which includes leap seconds.  To place timings on a system
free of leap seconds, we provide the transit timing measurements from
this study along with previous transit timing measurements from the
literature on the uniform Barycentric Julian Date Terrestrial Time
(TT) system in Table~\ref{tab:ttv}.  To refine the ephemeris of \xonb,
we analyze precise transit timing measurements from the refereed
literature based upon a complete transit and not impacted by strong
trends.  A subset of the transit timing measurements from
\citet{HOL06} and \citet{CAC09} meet these characteristics.
Specifically, the first 6 transit timing measurements listed in
Table~\ref{tab:ttv} are included.  The resulting ephemeris is given in
Table~\ref{tab:planet}, and the timing residuals from the linear
ephemeris model are shown in the top panel of Figure~\ref{fig:ttv1}.
With the published uncertainties on the transit timing measurements,
the linear ephemeris model is a moderately poor fit to the transit
times.  The resulting $\chi^2=16.08$ with $\nu=4$ degrees of freedom
has a 2.98 $\sigma$ chance of occurring randomly.  A second set of
uncertainties for the ephemeris of \xonb\ is given in
Table~\ref{tab:planet} that result from scaling the published transit
time uncertainties to enforce a reduced $\chi^{2}=1$ for a linear
ephemeris.  The significance of rejecting the linear ephemeris depends
most on transit event 86 in Figure~\ref{fig:ttv1}.  Without transit
event 86, the linear ephemeris fit improves to $\chi^2=6.94$ with
$\nu=3$, which has a 1.79 $\sigma$ of chance occurrence indicating an
acceptable fit.

Additional transit timing measurements of \xonb, with lower precision,
are consistent with the linear ephemeris determined from the
higher precision transit timings.  We select additional transit timing
measurements of \xonb\ from the Exoplanet Transit Database
\citep{POD10} and the Amateur Exoplanet Archive (AXA)\footnote{http://brucegary.net/AXA/x.htm} that have observations based upon $>$~half of the transit and
not impacted by strong trends.  We list the additional transit timings
that satisfy these characteristics in Table~\ref{tab:ttv}. The transit
timings are converted to the BJD (TT) timing system, and we also
include the measurement from \citet{RAE09}.  The bottom panel in
Figure~\ref{fig:ttv1} shows the residual timings (Black points) from
the linear ephemeris model, which was determined from the higher
quality transit timings (Red points).  These additional transit
timings have a $\chi^2=23.2$ for $\nu=23$ indicating the linear
ephemeris model is adequate for these data.

The transit timing residuals from a linear ephemeris shown in
Figure~\ref{fig:ttv1} qualitatively give the appearance for a
sinusoidal pattern (Gary, B., private communication).  Adding a long
period sinusoidal component to a linear ephemeris results in a
$\Delta\chi^{2}=11.6$ improvement from the linear ephemeris, with a
period for the sinusoidal component of 118.3 transit cycles (463 day).
Bayesian evidence is one method to quantify the most probable model
that describes the data.  The details to calculate the Bayesian odds
ratio between a linear ephemeris model and the linear+sinusoid
ephemeris is given in Appendix~\ref{apx:bay}.  Using the transit
timing measurements in Table~\ref{tab:ttv}, the odds ratio of the
linear+sinusoidal model to linear model is 5:1.  This is substantial
evidence in favor of the linear+sinusoidal model, but much less than
the 100:1 odds ratio typically accepted as a decisive result.

Overall, the linear+sinusoidal model cannot be distinguished from the
linear model, and given the current data both models are acceptable
descriptions of the data.  The most likely value for $P_{\rm
sin}=6.84$ transit events (26.77 day), with the previously identified
longer period $P_{\rm sin}=120$ transit events (470 day), the second
most likely solution.  The preference for a linear+sinusoid model is
weakened, odds ratio 2:1, when the transit timing measurement from the
top panel of Figure~\ref{fig:ttv1} that most significantly departs
from the linear ephemeris is not included in the analysis.

One potential source of Transit Timing Variations (TTV) is the
dynamical influence of additional planets orbiting \xon\
\citep{HOL05,AGO05}. In particular, the amplitude of TTV for Earth mass
planets orbiting at the 2:1 interior or exterior mean-motion resonance
with a hot Jupiter planet like \xonb\ can reach $>$50~s \citep{HAG09}.
A full analysis on the limits for companions to \xonb\ from the null
detection of TTV for \xonb\ is beyond the scope of this work
\citep{AGO05,AGO07}.  However, we use the Bayesian evidence
calculation to rule out the presence of Earth mass planets in an
exterior 2:1 mean motion resonance with \xonb.  We model the TTV of
\xonb\ due to a coplanar companion at the exterior 2:1 mean motion resonance by
adding a sinusoid with an expected period of 72.2$P$ \citep{AGO05} to
the transit timing measurements in Table~\ref{tab:ttv}.  Amplitudes
$>$45~s for the sinusoid TTV result in an odds ratio $>$100:1 for a
decisive detection.

\section{Discussion \& Conclusion}\label{sec:disc}

Without the interference of the Earth's atmosphere, space based
observations provide an opportunity to achieve high precision,
Poisson-limited photometric observations.  However, the thermal
forcing from the orbital cycle of HST and uncorrected instrumental
systematics present a challenge to Poisson limited performance.  We
correct for gain-like variations in high cadence, high precision time
series observations in NICMOS, which improve by a factor of two the
relative flux measurement.  Although the baseline of observations is
limited, there is some evidence that the non-repeatability in the G141
grism positioning between orbits affects the system throughput at the
$0.12\%$ level in relative flux.  When observations taken with similar
grism positioning are grouped together, only a few terms related to
the PSF positioning and shape are needed to decorrelate the
systematics that appear on the HST orbital time scale.  The remaining
source of noise is consistent with Poisson and the uncertainty in
determining the background level.  Despite the improved modeling of
the systematics, the physical transiting planet model contains some
degeneracy with the corrections for systematics.  The largest impact
is for determining the transit depth, $R_{p}/R_{\star}$, which has an
uncertainty 5 times larger than if the observations were of equivalent
quality but free from red-noise systematics.  Parameters such as
$a/R_{\star}$ and transit timing are less sensitive to the treatment
of red-noise systematics. \citet{SOU08} reaches similar conclusions
for the analysis of the high quality light curves of HD 209458b taken
with the STIS on HST \citep{BRO01B,KNU07}. This emphasizes the
necessity of simultaneous fitting of the transit model and systematics
in order to provide a more realistic assessment of the uncertainties.

Despite the limitations from needing to correct for the systematics,
we greatly improve upon the precision for determining the properties
of \xonb\ directly measurable from the light curve such as
$a/R_{\star}$, $R_{p}/R_{\star}$, $\rho_{\star}$, and transit timing.
We find no significant difference in the determination of
$R_{p}/R_{\star}$ between the NIR and previous determinations in the
optical \citep[][Reproduced in
Tables~\ref{tab:planet}~\&~\ref{tab:star}]{TOR08,SOU08}.  Using the
same data set as this study, \citet{TIN10} independently find a
similar absorption depth averaged over the full wavelength coverage.
Using transmission spectroscopy, precise and accurate measurements of
the transit depth as a function of wavelength are sensitive to opacity
variations in the upper levels (1 mbar) of the planetary atmosphere,
and theoretical planetary atmosphere models can be used to interpret
these measurements to constrain the abundance of molecules and
temperature-pressure profiles \citep{SEA00,BRO01,MAD09,FOR10}.  For
example, in their analysis of G141 grism data for HD 149026b,
\citet{CAR09B} find a larger absorption depth averaged over the G141
grism when compared to optical and Mid-IR measurements than the
theoretical models predict.  The technical improvements to NICMOS data
analysis that we outline in this study may lead to a more precise
estimate of the absorption depth in the large archive of NICMOS grism
observations of transiting extrasolar planets.

The transit light curve alone does not allow measuring $R_{p}$, and
with the quality of light curve from this work, the uncertainty in
$R_{p}$ is dominated by the adopted estimate for $M_{\star}$ based
upon stellar isochrones \citep{TOR08}.  Alternatively, measuring the
parallax of \xon, along with $T_{eff}$, apparent magnitude, and
bolometric correction determinations, yields a constraint on
$R_{\star}$.  This constraint on $R_{\star}$ and the transit light
curve will provide a separate estimate for $M_{\star}$.  As part of
the proposal for these NICMOS observations, Fine Guidance Sensor (FGS)
on HST observations were obtained to measure the parallax of \xon.  We
estimate the FGS distance will provide an estimate of $R_{\star}$ with
an uncertainty of 5\%, and this will provide a 15\% constraint on
$M_{\star}$ using the transit light curve from this study.
Asteroseismology provides an additional estimate of $\rho_{\star}$ and
has recently been demonstrated to yield very precise estimates for the
stellar host and planet in the case of HD 17156b (Nutzman, in
preparation), and will be a routine procedure for the transiting
planets found around the brighter stars with the Kepler mission
\citep{GIL10}.  Measuring the bulk mass and radius of planets with the
highest precision will require improvements in understanding the mass
and radius of their stellar hosts \citep{SOU08}.

The improved precision of the NICMOS time series photometry has
provided two additional measurements of the mid-transit time of
arrival.  The timing between the two HST visits analyzed in this work
is different by 2$\sigma$ from a linear ephemeris.  Also, a single
transit measurement from \citep{CAC09} departs most significantly from
a linear ephemeris.  Based upon the current measurements of \xonb\
transit timings, any bona fide TTV for \xonb\ will likely have a
peak-to-trough amplitude $<$90 s, and if the TTV are concentrated in
single events, then single timing precision needs to be $<$23~s.  The
HST observations presented in this study achieved $\sigma$=8.6~s and
$\sigma$=15~s precisions in transit timing.  The non continuous nature
of HST observations negatively impacts the transit timing.  The timing
precision with HST can be improved by optimizing coverage of the
ingress and egress portion of the light curve. However, transit timing
measurements with $\sigma$=13~s precision are possible with ground
based observations as demonstrated on \xonb\ by \citet{HOL06}.  More
detailed analysis is warranted, but the current transit timing data
rule out coplanar Earth mass companions to \xonb\ orbiting in 2:1 mean
motion resonance.  There have been no additional radial velocity
measurements published since the discovery paper
\citep{MCC06}.  Additional radial velocity measurements can aid in
constraining the presence of additional planets orbiting \xon.

The Bayesian evidence calculation that we outline in Appendix~\ref{apx:bay}
can be applied to other transit timing data sets.  The simplified
modeling of TTV from a linear ephemeris with a sinusoidal component
makes the model selection process numerically expedient.  As the
quantity and quality of transit timing measurements expands from
ground-based observations and especially with the contributions of the
\corot\ and \kepler\ missions \citep{CSI10,GIL10B}, early discovery of TTV
will aid in prioritizing and planning the followup observations \citep{FOR08}.

\acknowledgments

This work benefited from discussions with Philip Nutzman, Dan
Fabrycky, Joe Hora, David Charbonneau, and Jeff Stys.  We thank Mark
Swain and Pieter Dieroo for conversations about spectrophotometry with
NICMOS.  We thank Josh Winn and Valentin Ivanov for discussing their
published transit timings, and the following observers for their
generosity in making their transit timing measurements publicly
available, Anthony Ayiomamitis, Cindy Foote, Bruce Gary, Joao
Gregorio, Bill Norby, Gregor Srdoc, Jaroslav Trnka, and Tonny
Vanmunster.

\begin{appendix}
\section{Bayesian Evidence for Sinusoidal Transit Timing Variations}\label{apx:bay}
Bayesian evidence is one method to quantify the most probable model
that describes the data, and we follow the procedure as outlined in
\citet{GRE05,GRE05B,GRE07,FOR07} to calculate the Bayesian odds ratio
between the linear ephemeris model and the linear+sinusoid ephemeris
model.  The algorithm presented by \citet{GRE05} is tailored to
quantify the odds for a radial velocity data set to be modeled by no
planet, one planet, or $>$~one planet, and we independently implement
the algorithm for the model comparison problem at hand.  We model the
transit timing ephemeris as
\begin{equation}
T_{o}+E\times P_{\rm lin}+a\sin\left( \frac{2\pi E}{P_{\rm sin}}+\phi\right) ,\label{eq:ttvmod}
\end{equation}
where the free parameters are the ephemeris zeropoint, $T_{o}$, linear
ephemeris period, $P_{\rm lin}$, amplitude of the sinusoidal
component, $a$, sinusoidal component period, $P_{\rm sin}$, and
sinusoidal component phase, $\phi$.  The independent variable in the
model is the transit event, $E$.  Thus, we calculate the odds ratio
between the linear ephemeris model with only two free parameters,
$T_{o}$ \& $P_{\rm lin}$, and the full linear+sinusoid model with five
free parameters as shown in Equation~\ref{eq:ttvmod}.  We employ the
independent Gaussian model for the residuals in the likelihood.  We
include an additional free parameter in the likelihood that scales the
reported errors on the transit timing measurements, $s$.  Allowing $s$
to vary enables the odds ratio between the competing models to take
into account our incomplete knowledge as to the overall scaling of the
transit timing uncertainties.  However, we are assuming that the
transit timing uncertainties are correct in a relative sense.
\citet{FOR07B} employ the same linear+sinusoid ephemeris model to
characterize and simulate the detection of the TTV signature from a
Trojan companion using a lomb-scargle periodogram approach.

The priors for $T_{o}$ and $P_{\rm lin}$ are uniform since they are
already well determined, and their prior range accommodates the
constraints placed by the data.  We adopt a Jeffreys prior for $s$
that is constrained between 0.25$<s<$4.0.  We adopt a modified
Jeffreys prior for $a$ with a break at 1.0~s and maximum of $10^4$~s.
The maximum $a$ is the largest expected amplitude of transit timing
variations expected in the planetary regime \citep{HOL05}.  Based upon
the discussion in \citet{GRE07}, we choose a Jeffreys prior for
$P_{\rm sin}$ with a minimum of 0.9~transit events and a maximum of
1000~yr.  The prior for $0<\phi <2\pi$ is uniform.  The PT MCMC
algorithm has 7 parallel chains, and an automated iterative algorithm
of proposal step size adjustments until the acceptance fraction,
$0.2<f<0.3$, is reached. The calculation is based upon the MCMC steps
after the proposal step sizes are finalized.  The linear+sinusoid
ephemeris model is not optimized for the complicated signature TTV
signals that can be present and will underestimate the importance of a
TTV signature that is concentrated in few events.  However, a variety
of TTV signatures are qualitatively sinusoidal
\citep[e.g.,][]{FOR07B,HAG09}, and in general, the dynamical variations of
the osculating elements in resonant interactions are well described by
expansions of oscillating terms \citep{MUR99}.

The Bayesian Information Criterion (BIC) is an approximation to the
Bayesian evidence that is most accurate in the limit of a large number
of measurements with a simple likelihood surface \citep{KAS95}.  In
this example, the number of transit timing measurements is small, and
the multimodal likelihood surface invalidates the Taylor series
expansion approximation implicit in BIC.  However, the
$\Delta$BIC=5.2, between the linear and linear+sinusoidal model mildly
prefers the linear+sinusoidal model.  $\Delta$BIC$>$10 is necessary
for a decisive decision with the BIC
\citep{KAS95}.  Both the Bayesian evidence and BIC show that there is
a mild preference for the linear+sinusoidal model, but not a definitive
one. 
\end{appendix}

\clearpage

\begin{figure}
\plotone{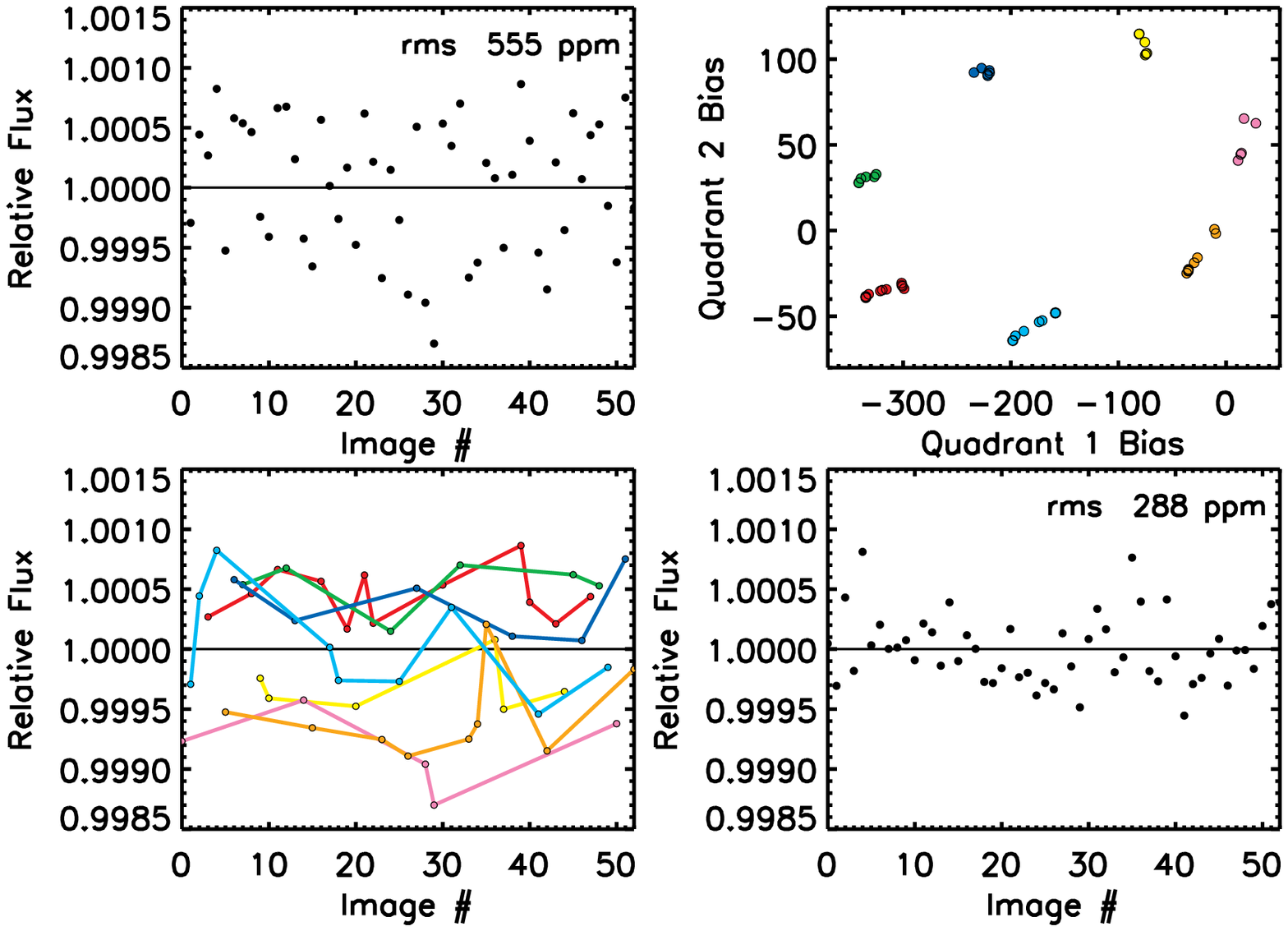}
\caption{
{\it Top Left:} Raw relative flux for the single orbit fully in
transit as a function of image number within an orbit.  The noise is
2.2 times higher than expected and broadly distributed
(i.e., non-Gaussian).  {\it Top Right:} Zeroth-read bias level from the
lower left detector quadrant (quad 1) versus the the lower right detector
quadrant (quad 2) for the same orbit as shown in top left figure.  The points
are color coded into 7 distinct state groupings. {\it Lower Left:}
The relative flux observations that share the same 7 state grouping
are connected by lines with the same color coding as the top right
figure.  The observations for each state are systematically above or
below the mean flux level.  {\it Lower Right:} Each group is forced to the
average flux level of an orbit resulting in nearly a factor of 2
improvement in the noise.  The remaining correlated red-noise is
removed by decorrelation with external parameters.\label{fig:7cor}}

\end{figure}

\begin{figure}
\plotone{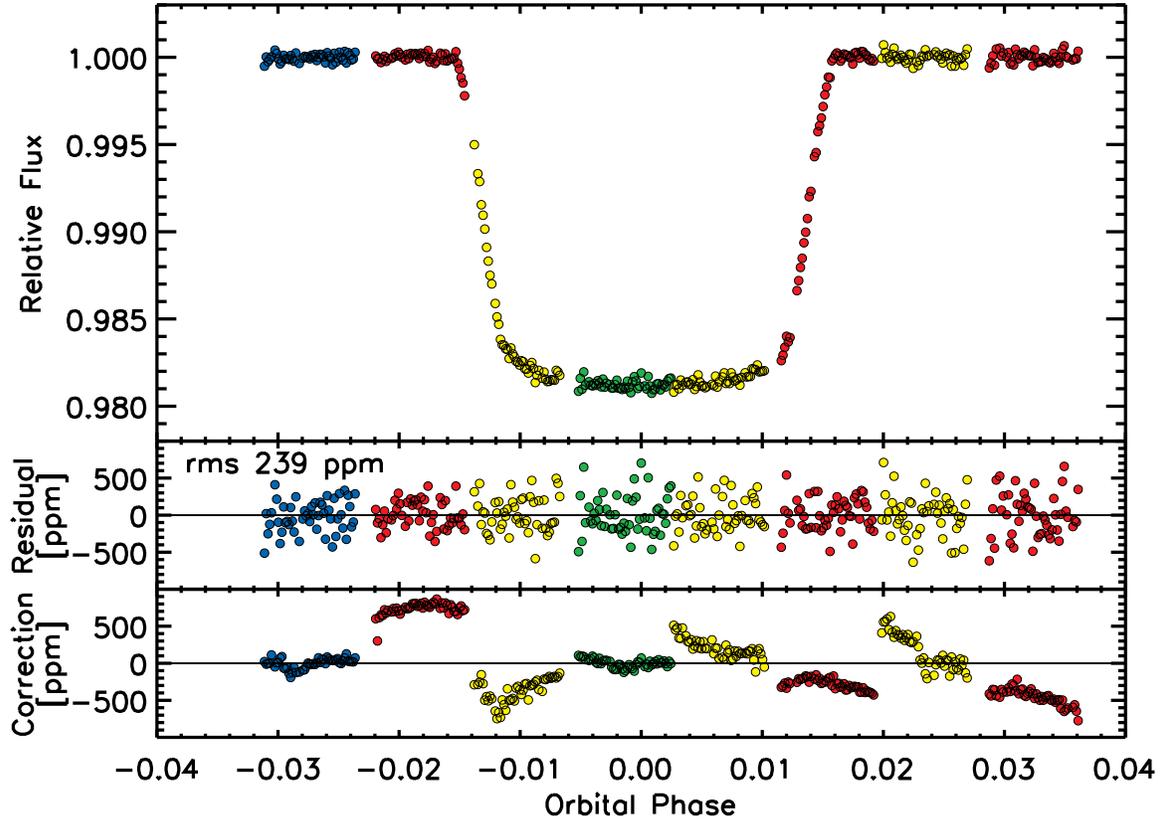}
\caption{
{\it Top:} Best fit relative flux of the observations after correction
for the 7 readout states and decorrelation with external parameters.
{\it Middle:} Residual from the best fit transit model and
corrections.  {\it Bottom:} Best fit correction for the external
parameter decorrelation.  The color coding is the same as Figure~\ref{fig:rawlc}\label{fig:finlc}}
\end{figure}

\begin{deluxetable}{lccc}
\tabletypesize{\footnotesize}
\tablewidth{0pt}
\tablecaption{{\rm Planet Properties} - \xonb\ }
\startdata
\hline
\hline
Parameter                                       & This Work         & \citet{TOR08}   & \citet{SOU08,SOU09}\\
\hline
Mass, $M_{\rm p}$ \brk{\Mjup} 			& \vMp$\pm$\eMp\    & 0.92$\pm$0.08  & 0.94$\pm$0.07 \\
Radius, $R_{\rm p}$ \brk{\Rjup} 		& \vRp$\pm$\eRp\    & 1.21$\pm$0.04 & 1.22$\pm$0.07 \\
Velocity semiamplitude, $K$ \brk{\mps} 	& \vrvK$\pm$\ervK\tablenotemark{c} & \nodata  & \nodata \\
Semimajor axis, $a$ \brk{A.U.} 			& \vap$\pm$\eap     & 0.0493$\pm$0.0009 & 0.0499$\pm$0.0008 \\
Orbital inclination, $i $ \brk{deg} 		& \vincl$\pm$\eincl & 88.8$\pm^{+0.7}_{-0.3}$ & 89.1$\pm$0.8 \\
Scaled semimajor axis, $a/R_{\star}$            & \vadr$\pm$\eadr   & 11.55$\pm^{+0.03}_{-0.45}$ &  \nodata \\
Planet-to-star radius ratio, $R_{\rm p}/R_{\star}$ & \vrho$\pm$\erho & 0.1326$\pm$0.0004 & 0.1317$\pm$0.0019 \\
Transit duration (1$^{\rm st}$-4$^{\rm th}$ contact), $\tau_{I-IV}$ \brk{hr} & \vdur$\pm$\edur & \nodata & \nodata \\
Ingress/Egress duration (1$^{\rm st}$-2$^{\rm nd}$ contact), $\tau_{I-II}$ \brk{hr} & \ving$\pm$\eing & \nodata & \nodata \\
Impact parameter, b                             & \vbimp$\pm$\ebimp & 0.24$\pm^{+0.04}_{-0.14}$ & \nodata \\
Planet-to-orbit radius ratio, $R_{\rm p}/a$     & \vrpda$\pm$\erpda & \nodata         & 0.01166$\pm$0.00035 \\
Star-to-orbit radius ratio, $R_{\star}/a$       & \vrsda$\pm$\ersda & \nodata         & 0.0886$\pm$0.0019 \\
Summed radius-to-orbit ratio, $(R_{\star}+R_{\rm p})/a$ & \vrsrp$\pm$\ersrp & \nodata & 0.1003$\pm$0.0022 \\
Planet gravity, log$g_{\rm p}$ \brk{cgs}        & \vlggp$\pm$\elggp\ & 3.21$\pm$0.04 & 3.199$\pm$0.04 \\
Planet density, $\rho_{\rm p}$ \brk{\gpcmthree} & \vdensp$\pm$\edensp & 0.65$\pm$0.09 & 0.69$\pm$0.08 \\
Safronov, $\Theta$                              & \vsaf$\pm$\esaf   & 0.0744$\pm$0.006 & \nodata \\
\hline
Orbital Period, $P$ \brk{day} & \multicolumn{3}{l}{\vperiod\ $\pm$ \eperiod\ $\pm$ 0.0000018\tablenotemark{a}} \\
Time of midtransit, $T_o$ \brk{BJD(TT)}	& \multicolumn{3}{l}{ \vjd\ $\pm$ \ejd\ $\pm$ 0.00022\tablenotemark{a,b}} \\
\enddata
\tablenotetext{a}{The second set of uncertainties correspond to enforcing reduced $\chi^2=1$ for a linear ephemeris model through scaling the uncertainties in the transit timing measurements.}
\tablenotetext{b}{Time is on the BJD Terrestrial Time (TT) system.  For times on the BJD (UTC) system subtract 65.184~s.}
\tablenotetext{c}{Adopted from \citet{MCC06}}
\label{tab:planet}
\end{deluxetable}

\begin{deluxetable}{lccc}
\tabletypesize{\small}
\tablewidth{0pt}
\tablecaption{{\rm Stellar Properties} - \xon\ }
\startdata
\hline
\hline
Parameter                                & This Work                            & \citet{TOR08}                & \citet{SOU09} \\
\hline
Mass, $M_{\star}$ \brk{$M_{\odot}$}      & \vMs$\pm$\eMs\tablenotemark{a}       & \nodata                      & 1.066$\pm$0.051 \\
Radius, $R_{\star}$ \brk{$R_{\odot}$}    & \vRs$\pm$\eRs\                       & 0.934$\pm$0.035              & 0.95$\pm$0.025 \\
Effective temperature, $T_{eff}$ \brk{K} & \vTeff$\pm$\eTeff\tablenotemark{a,b} & \nodata                      & \nodata \\
Metallicity \verb|[|Fe/H\verb|]|         & \vFeH$\pm$\eFeH\tablenotemark{a,b}   & \nodata                      & \nodata \\
Stellar gravity, log$g$ \brk{cgs}        & \vgs$\pm$\egs\                       & 4.509$\pm^{+0.018}_{-0.027}$ & 4.510$\pm$0.018 \\
Stellar density, $\rho_{\star}$ \brk{\gpcmthree} & \vdens$\pm$\edens\           & 1.877$\pm^{+0.015}_{-0.21}$  & 1.75$\pm$0.11 \\
First limb darkening coeff., $u_{1}$             &  $<_{1\sigma\, U.L.}$\eUone  & \nodata                      & \nodata \\
Second limb darkening coeff., $u_{2}$            &  \vUtwo$\pm$\eUtwo           & \nodata                      & \nodata\\
\enddata
\tablenotetext{a}{Adopted from \citet{TOR08}}
\tablenotetext{b}{Adopted from \citet{MCC06}}
\label{tab:star}
\end{deluxetable}

\begin{deluxetable}{lccc}
\tabletypesize{\small}
\tablewidth{0pt}
\tablecaption{{\rm Precision of Light Curve}}
\startdata
\hline
\hline
Parameter                               &  {\rm limb free}  & {\rm limb fixed}  & {\rm Sys fixed} \\
$a/R_{\star}$                           & 11.24$\pm$0.09    & 11.28$\pm$0.09    & 11.26$\pm$0.07  \\
$R_{\rm p}/R_{\star}$                   & 0.1320$\pm$0.0005 & 0.1316$\pm$0.0003 & 0.1316$\pm$0.0001 \\
$\tau_{I-IV}$ \verb|[|dy\verb|]|        & 0.1238$\pm$0.0003 & 0.1240$\pm$0.0002 & 0.1240$\pm$0.0002 \\
$t_{1}$ \verb|[|s\verb|]|               & $\sigma$=8.6      & $\sigma$=8.6      & $\sigma$=6.9      \\
\hline
\enddata
\label{tab:errors}
\end{deluxetable}

\begin{deluxetable}{lccc}
\tabletypesize{\small}
\tablewidth{0pt}
\tablecaption{{\rm Transit Timing Measurements}}
\startdata
\hline
\hline
BJDo (TT)\tablenotemark{a}  & $\sigma$ [day] & Source \\
\hline
\vjdOne & \ejdOne & This Work \\
\vjdTwo & \ejdTwo & This Work \\
\hline
2453883.80639 & 0.00018 & \citet{HOL06} \\
2453887.74753 & 0.00015 & \citet{HOL06} \\
2454222.77613 & 0.00036 & \citet{CAC09} \\
2454226.71843 & 0.00037 & \citet{CAC09} \\
\hline
2453808.9170 &  0.0010 & AXA\tablenotemark{b}; \citet{POD10} \\
2453832.5683 &  0.0020 & '' \\
2453879.8643 &  0.0011 & '' \\
2453887.7477 &  0.0006 & '' \\
2453903.5141 &  0.0008 & '' \\
2454214.8927 &  0.0010 & '' \\
2454285.8403 &  0.0010 & '' \\
2454553.8624 &  0.0010 & '' \\
2454620.8655 &  0.0015 & '' \\
2454620.8678 &  0.0008 & '' \\
2454624.8100 &  0.0014 & '' \\
2454628.7515 &  0.0004 & '' \\
2454648.4586 &  0.0011 & '' \\
2454888.8900 &  0.0007 & '' \\
2454908.5962 &  0.0010 & '' \\
2454959.8375 &  0.0005 & '' \\
2454963.7789 &  0.0005 & '' \\
2454963.7790 &  0.0010 & '' \\
2454967.7192 &  0.0007 & '' \\
2454971.6613 &  0.0006 & '' \\
2454983.4866 &  0.0008 & '' \\
2454987.4285 &  0.0008 & '' \\
2455058.3769 &  0.0010 & '' \\
2455290.9235 &  0.0006 & '' \\
2554171.5333 &  0.0017 & \citet{RAE09} \\
\hline
\enddata
\tablenotetext{a}{NOTE: Times are on the BJD Terrestrial Time (TT) system.  For times on the BJD UTC system subtract 65.184~s for times before 2454832.5 Julian Date (UTC) and 66.184~s for times after.}
\tablenotetext{b}{Amateur Exoplanet Archive - http://brucegary.net/AXA/x.htm}
\label{tab:ttv}
\end{deluxetable}

\end{document}